\documentclass[a4paper,10pt]{article}
\usepackage[utf8x]{inputenc}

\title{ Non-anticommutative ABJ Theory  }
\author{ Mir Faizal \\
Mathematical Institute, University of Oxford
\\ Oxford
OX1 3LB, United Kingdom 
 }

\begin{document}

\maketitle

\begin{abstract}
In this paper we  will discuss  non-anticommutative 
deformations of  the harmonic superspace. We will analyse 
the non-anticommutative  deformation of the  superspace that
break the supersymmetry from  $\mathcal{N} =3$ supersymmetry to 
$\mathcal{N} =2$ supersymmetry. We will then study the ABJ theory
in this non-anticommutative superspace. This deformed ABJ theory 
will be shown to posses $\mathcal{N} =5$ supersymmetry. 
\end{abstract}

\section{Introduction}
In four dimensions  $\mathcal{N} =2$ supersymmetry has been studied in harmonic superspace  \cite{h1, h2}, and this has been adopted for  
analysing $\mathcal{N}= 3$ supersymmetry in three dimensions \cite{h21, h3, h4}. The harmonic superspace variable parameterize the coset $SU(2)/U(1)$ and are
 well suited for analysing theories with hight amount of supersymmetry. Thus, harmonic superspace 
 has been used for analysing the ABJM theory \cite{ahs}, which is a superconformal Chern-Simons-matter theory with manifest
 $\mathcal{N} =6$ supersymmetry \cite{1, 2, 3, 4}. 
This theory is thought to be a low energy description of $N$ M2-branes  on $C^4/Z_k$ orbifold because it coincides with the BLG theory 
for the only known example of a Lie $3$-algebra \cite{BL1, BL2, BL3, blG}. So, the supersymmetry of the ABJM theory is  expected to be 
enhanced to full $\mathcal{N} =8$ supersymmetry for $k =1,2$ \cite{sabjm, sabjm1}. 
The gauge fields in the ABJM theory are governed by the Chern-Simons 
action and the matter fields live in the bifundamental representation of the gauge group $U(N) \times U(N)$. 
A generalization of the ABJM theory to the case 
where the matter fields live in the bifundamental representation of gauge group $U(M) \times U(N)$ with $M \neq N$ has been made  \cite{5, 5ba, 5a1, 5a2}.
This theory is called 
the ABJ theory and it also has $\mathcal{N} =6$ supersymmetry. 
However, unlike the ABJM theory,  non-planar corrections to the two-loop dilatation generator of ABJ theory
 mix states with positive and negative parity, and this mixing is proportional to $M − N$ \cite{5a}. So, for $M =N$, when the 
ABJ theory reduces to the ABJM theory, there is no mixing.

In string theory the $NS$  background causes a noncommutative  deformation between the spacetime 
coordinates   
 \cite{sw, dn, dfr, co}, and a gravitino background causes a noncommutative deformation between 
the spacetime and Grassmann coordinates 
 \cite{bgn, gp,  gp01, gp1}. All these deformations preserve the 
supersymmetry of the theory. However, the presence of a $RR$ background causes a deformation 
between the Grassmann  coordinates and thus  breaks 
the a certain amount of the supersymmetry of the theory 
\cite{ov,se,beta1,beta2,beta3,beta4}. In four dimensions this can give rise to a fractional 
amount of supersymmetry  like $\mathcal{N} =1/2$ supersymmetry. 
Non-anticommutative deformation of harmonic superspace has also been 
analysed \cite{8l,  82, 83, 84}.

As M-theory is dual to type II string theory, a deformation of the string theory side will also
 generate a deformation on the M-theory side. 
Thus, a noncommutative deformation of the 
M-theory will be dual to a noncommutative 
deformation of type II string theory caused by $NS$ background. 
Similarly, a non-anticommutative deformation of the M-theory will 
be dual to a non-anticommutative deformation of type II string theory caused by 
$RR$ background. In fact, this duality can be explicitly verified by using the
 novel Higgs mechanism \cite{zz1,zz2,zz3,zz4a}. Thus, if we perform the higgsing of
 non-anticommutative M2-branes we will obtain non-anticommutative D2-branes. 
So, the non-anticommutative ABJ theory is dual to type II string theory deformed
 by a $RR$ background. 
Noncommutative  deformation of the M2-branes in $\mathcal{N} =1$ superspace 
have been already studied \cite{4, hi}. However,  non-anticommutative deformation of
the M2-branes has not been studied. 

This non-anticommutative deformation  of the 
M2-branes can occur due to the presence of a curved three form field $C_{\mu\nu\tau}$.
This is  because the ABJM theory is the  
 boundary gauge theory dual to the eleven dimensional supergravity on $ AdS_{4}\times S_7/ Z_k$.
 A deformation of this eleven dimensional supergravity on $ AdS_{4}\times S_7/ Z_k$ can be 
caused by a three form field $C_{\mu\nu\tau}$. A constant three form field $C_{\mu\nu\tau}$ 
is only expected to change the gauge group of the theory without breaking 
any supersymmetry. Thus, the ABJ theory can be viewed as a deformation of the ABJM theory. 
However, a deformation of the eleven dimensional supergravity on $ AdS_{4}\times S_7/ Z_k$
 by a curved  
three form field $C_{\mu\nu\tau}$ will change the geometry considerably and is expected 
to partially break 
the supersymmetry. The boundary theory 
dual to this deformed eleven dimensional supergravity will be a non-anticommutative 
ABJ theory. 

Now, if $H_{\mu\nu\tau\rho}$ the field strength of 
this three form field, then  we expect a non-anticommutative 
deformation proportional to  $ \{ {\theta_a ,\theta_b} \} \sim
 (\gamma^\mu\gamma^\nu\gamma^\tau\gamma^\rho H_{\mu\nu\tau\rho})_{ab}  $ to occur.
However, it 
was not possible to study the non-anticommutative deformations of the ABJM theory in 
the $\mathcal{N} =1$ superspace as that 
would break all the manifest supersymmetry of the superspace. 
It would thus be interesting to analyse the non-anticommutative deformations of this
theory  in superspace with higher amount of manifest supersymmetry. So, in this paper 
we analyse the non-anticommutative 
deformation of the ABJ theory in harmonic superspace. Non-anticommutative deformations  break
  the total supersymmetry of this theory from $\mathcal{N} =6$ supersymmetry to
 $\mathcal{N} =5$ supersymmetry.

\section{Harmonic superspace}
In this section we will review harmonic superspace in three dimensions \cite{h21, h3, h4, h5}. 
These harmonic variable are parameterize by the coset $SU(2)/U(1)$. 
So, the
harmonic variables $u^{\pm}$ are subjected to the 
constraints 
$
 u^{+i} u^-_i = 1, \,\, u^{+i} u^+_i = u^{-i} u^-_i =0,  
$
and the superspace coordinates are given by
\begin{equation}
 z = ( x^{ab}, \theta_{a}^{++}, \theta^{--}_a, \theta^0_a, u_i^{\pm} ),
\end{equation}
where $ \theta^{\pm}_a = \theta_a^{ij} u^{\pm}_i u^{\pm}_j$ and $\theta^0_a = \theta^{ij}_a u^+_i u^-_j$. 
In the harmonic superspace  the following derivatives are constructed 
\begin{eqnarray}
&\partial^{++}=u^+_i\frac\partial{\partial u^-_i}, &
\partial^{--}=u^-_i\frac\partial{\partial u^+_i},\nonumber \\ 
&\partial^0 = u^+_i\frac\partial{\partial u^+_i}
-u^-_i\frac\partial{\partial u^-_i}.& 
\end{eqnarray}
These derivatives are used to define the following derivatives
\begin{eqnarray}
D^{--}_a=\frac\partial{\partial\theta^{++ a}}
 +2i\theta^{--b}\partial^A_{ab}, &&
D^0_a= -\frac12\frac\partial{\partial\theta^{0 a}}
+i\theta^{0 b}\partial^A_{ab},\nonumber \\ 
D^{++}_a=\frac{\partial}{\partial
\theta^{--a}}.
\end{eqnarray}
Apart from these derivatives, the following derivatives are also constructed
\begin{eqnarray}
\nonumber \\ 
{\cal
D}^{++}&=&\partial^{++}+2i\theta^{++ a}\theta^{0 b}
 \partial^A_{ab}
 +\theta^{++a}\frac\partial{\partial\theta^{0 a}}
 +2\theta^{0 a}\frac\partial{\partial\theta^{--a}},\nonumber \\
{  D}^{--}&=&\partial^{--}
 -2i\theta^{--a}\theta^{0 b}\partial^A_{ab}
 +\theta^{--a}\frac\partial{\partial\theta^{0 a}}
 +2\theta^{0 a}\frac\partial{\partial\theta^{++ a}},
\nonumber \\
{  D}^0&=&\partial^0+2\theta^{++ a}\frac\partial{\partial\theta^{++ a}}
-2\theta^{--a}\frac\partial{\partial\theta^{--a}}.
\end{eqnarray}
The superalgebra  satisfied by these derivatives is given by 
\begin{eqnarray}
 \{D^{++}_a, D^{--}_b\}=2i\partial^A_{ab}, \quad \{D^{0}_a,
D^{0}_b\}=-i\partial^A_{ab}, 
\nonumber \\ 
{[{ D}^{\mp\mp}, D^{\pm\pm}_a]}=2D^0_a, \quad [{  D}^{0},
D^{\pm\pm}_a]=\pm 2D^{\pm\pm}_a, 
\nonumber \\ 
\partial^0=[\partial^{++},\partial^{--}],\quad
[{  D}^{++}, {  D}^{--}]={  D}^0. \nonumber \\
\{D^{\pm\pm}_a, D^{0}_b\} = 0\,,\quad [{\cal
D}^{\pm\pm}, D^0_a]=D^{\pm\pm}_a.
  \end{eqnarray}

The harmonic superspace has $\mathcal{N} =3$ supersymmetry in three dimensions. 
The generators of this $\mathcal{N} =3$ supersymmetry in three dimensions are given by 
\begin{eqnarray}
&Q^{++}_a=u^+_iu^+_j Q_a ^{ij}, &
Q^{--}_a=u^-_iu^-_j Q_a ^{ij},\nonumber \\ 
&Q^0_a = u^+_iu^-_j Q_a ^{ij},& 
\end{eqnarray}
where 
\begin{equation}
 Q_a^{ij} = \frac{\partial}{\partial \theta^a_{ij}} - \theta^{ijb }  \partial_{ab}.
\end{equation}
In harmonic superspace, the superfields which are independent of the
$\theta^{--}_a$ are called analytic
superfields. Thus, these analytic superfields satisfy
  \begin{eqnarray}
   D^{++}_a\Phi_A=0 \quad \Rightarrow \quad \Phi_A = \Phi_A(\zeta_A),
  \end{eqnarray}
where the
 coordinates parameterizing the analytic subspace 
are given by
\begin{eqnarray}
\zeta_A=(x^{ab}_A,
\theta^{++}_a, \theta^{0}_a, u^\pm_i).
  \end{eqnarray}
Here $x^{ab}_A$ is given by 
\begin{eqnarray}
x^{ab}_A=(\gamma_m)^{ab}x^m_A=x^{ab}
+i(\theta^{++a}\theta^{--b}+\theta^{++b}\theta^{--a}).
  \end{eqnarray}
It is convenient to define 
  \begin{eqnarray}
d^9z &=&-\frac1{16}d^3x
(D^{++})^2 (D^{--})^2(D^{0})^2, \nonumber \\
d\zeta^{(-4)}&=&\frac{1}{4} d^3x_Adu (D^{--})^2(D^{0})^2\,.
  \end{eqnarray}
Furthermore, a conjugation in this  superspace is defined  by
  \begin{eqnarray}
\widetilde{(u^\pm_i)}=u^{\pm i},\quad \widetilde{(x^m_A)}=x^m_A, \nonumber \\  \widetilde{(\theta^{\pm\pm}_a)}=
\theta^{\pm\pm}_a,\quad \widetilde{(\theta^0_a)}=
\theta^0_a.
  \end{eqnarray}
The analytic superspace measure is real
 $\widetilde{d\zeta^{(-4)}}=d\zeta^{(-4)}$
and the full superspace measure is imaginary $\widetilde{d^9z}=-d^9z$
because this conjugation  is squared to $-1$ on the harmonics  and to $1$ on $x^m_A$  and 
 Grassmann coordinates.  
\section{Deformation of Harmonic Superspace}
In this section we will analyse non-anticommutative deformation of the Harmonic superspace. Such
deformation occurs due to a
$RR$ background  in the string theory \cite{ov,se}. 
Unlike the noncommutative deformation caused by a $NS$  background or gravitino background, the non-anticommutative deformation 
a certain amount of the
breaks the supersymmetry of the theory. In four dimensions this deformation can be used to break half the supersymmetry of a  
$N = 1$ supersymmetry theory to obtain a theory with $N =1/2$ supersymmetry. This is because 
in four dimensions   the supersymmetric 
generator $Q_A$  can be split into $Q_a$ and $Q_{\dot{a}}$. So, it is possible to break the supersymmetry corresponding to 
one of these generators without breaking  the other one. This is thus done by imposing the following anticommutator
\begin{equation}
 \{\hat \theta_a, \hat\theta_b \} =C_{ab}. 
\end{equation}
This will break the supersymmetry corresponding to $Q_a$ without breaking the supersymmetry corresponding to $Q_{\dot{a}}$. 
We could also break the supersymmetry corresponding to $Q_{\dot{a}}$ without breaking breaking the supersymmetry corresponding to $Q_a$.
Now if we project the undeformed  $\mathcal{N} =1$ supersymmetric in four dimensions to three dimensions, then it will have 
 $\mathcal{N}= 2$ supersymmetry. This is because in three dimensions both $Q_a$ and $Q_{\dot{a}}$ act as separate supercharges. Now the 
$\mathcal{N} =1/2$ supersymmetry in four dimensions corresponded to $\mathcal{N} = (1, 0)$ or $\mathcal{N} = (0, 1)$  in three dimensions.
 It is not possible to obtain a $N =1/2$ supersymmetry in three dimensions from this as there are enough degrees of freedom to do so.
 
 The  $\mathcal{N} =2$ supersymmetric theory in four dimensions is 
generated by four supercharges. It is possible to break the supersymmetry with respect to any number of these supercharges. 
Thus, if the 
supercharges supercharges are denoted by $Q^{\pm}_a$ and and $Q_{\dot{a}}$. Then it is possible to obtain a $\mathcal{N} = 1/2$ theory 
by breaking the supersymmetry with respect to three of these supercharges by imposing the following anti-commutators
\begin{eqnarray}
  {\{ \hat \theta^+_{\dot{a}},\hat \theta^+_{\dot{b}} \} }=C^+_{\dot{a}\dot{a}}, &&   \{ \hat \theta^-_{\dot{a}}, \hat \theta^+_{\dot{a}} \} =C^-_{\dot{a}\dot{a}}, 
\nonumber \\
  {\{\hat \theta^+_a , \hat \theta^+_b \}} =C^+_{ab}.
\end{eqnarray}
From a three dimensional perspective this corresponded to breaking a $\mathcal{N} = 4$ supersymmetric theory to 
 $\mathcal{N} = (( 1, 0),( 0, 0))$. We could  obtain similar deformations, 
    $\mathcal{N} = ( (0, 1),( 0, 0))$,  $\mathcal{N} = ( (0, 0),( 1, 0))$ and 
 $\mathcal{N} = ( (0, 0), (0, 1))$, depending upon which supercharges are left undeformed. It is also be possible to deform the four dimensional
theory with  $\mathcal{N} =2$ supersymmetry 
as 
\begin{eqnarray}
  \{ \hat\theta^+_{\dot{a}}, \hat\theta^+_{\dot{b}} \} =C^+_{\dot{a}\dot{a}}, && 
  \{ \hat \theta^+_a, \hat \theta^+_b\} =C^+_{ab}.
\end{eqnarray} 
This will corresponding to  $\mathcal{N} = (( 1, 0),( 1, 0))$ in three dimensions. Now we can use generate other similar deformations 
 like $\mathcal{N} = (( 0, 1),( 1, 0))$, $\mathcal{N} = (( 0, 1),( 0, 1))$, $\mathcal{N} = (( 1, 0),( 0, 1))$. Finally, we can break only one of 
the supercharges in the four dimensional theory and obtain a $\mathcal{N} =2/3$ supersymmetric theory. Thus, if we impose 
\begin{eqnarray}
  \{ \hat \theta^+_{\dot{a}}, \hat \theta^+_{\dot{b}} \} =C^+_{\dot{a}\dot{a}}, 
\end{eqnarray} 
then we obtain a $\mathcal{N} =2/3$ in four dimensions. This corresponded to $\mathcal{N} = ((1,0), (1,1))$ in three dimensions. Similarly, we can 
obtain $\mathcal{N} = ((0,1), (1,1))$, $\mathcal{N} = ((1,1), (0,1))$ and $\mathcal{N} = ((1,1), (1,0))$,  in three dimensions. 

We could also start from the harmonic superspace in three dimensions and impose the following deformation 
\begin{equation}
 \{\hat \theta^{++}_a, \hat \theta^{++}_a \} = C^{++}_{ab}.
\end{equation}
This will break the supersymmetry corresponding to $Q^{++}_a$ without breaking  the supersymmetry corresponding to $Q^{--}_a$ and $Q^0_a$.
Thus, we will obtain $\mathcal{N} =2$ supersymmetry in three dimensions.  
This we could have similarly broken the supersymmetry with respect to $Q^{--}_a$ or $Q^0_a$ and left the remaining two intact
to obtain  $\mathcal{N} =2$ supersymmetry in three dimensions. We can also break 
the supersymmetry with respect to any two supercharges say $Q^{++}_a$ and $Q^{--}_a$ and leave the supersymmetry with respect to $Q^0_a$ intact by 
imposing the  following deformations 
\begin{equation}
 \{\hat \theta^{++}_a,\hat  \theta^{++}_a \} = C^{++}_{ab}, \quad \{\hat \theta^{--}_a, \hat \theta^{--}_a \} = C^{--}_{ab}.
\end{equation}
This way we will obtain a theory with $\mathcal{N} =1$ supersymmetric. 
We could have similarly have left either $Q^{++}_a$ or $Q^{--}_a$ intact to obtain a theory with $\mathcal{N} =1$ supersymmetric. 
It may be noted that $\mathcal{N} =3$ supersymmetry in three dimensions corresponded to $\mathcal{N} =3$ supersymmetry in two dimensions. 
This is because each of the supercharges splits into two independent supercharges $Q^{++}_{\pm}$, $Q^{--}_{\pm}$ or $Q^0_{\pm}$. Thus, 
 $\mathcal{N} =2$ supersymmetry in three dimensions will correspond to one of the following,  $ ((1,1),(1,1),(0,0))$, $((0,0),(1,1),(1,1))$
and $((1,1),(0,0),(1,1))$, in two dimensions. Similarly,  $\mathcal{N} =1$ supersymmetry in three dimensions 
will correspond to one of the following,  $ ((1,1),(0,0),(0,0))$, $((0,0),(1,1),(0,0))$
and $((0,0),(0,0),(1,1))$, in two dimensions. 
\section{Deformed ABJ Theory} 
In this section we will analyse non-anticommutative deformation of the ABJ theory in the harmonic superspace with one of generators of the
supersymmetry broken due to the deformation. Other deformations can be analysed in a similar way. 
So, to start with 
defining a vector field  $V^{++}$ in the harmonic superspace. 
We now deform the harmonic superspace 
by breaking the supersymmetry generated by $Q_a^{++}$ by imposing the following relations, 
\begin{equation}
 \{\hat \theta^{++}_a, \hat \theta^{++}_a \} = C^{++}_{ab}. \label{p} 
\end{equation}  
We now use Weyl
ordering and  express the Fourier transformation of a superfield on this deformed superspace as, 
\begin{eqnarray}
 \hat V^{++} ( \hat z ) = \int dp  V^{++}(p) \exp ( i p\hat z),   
\end{eqnarray}
where 
\begin{eqnarray}
 \exp (i p\hat z) &=& \exp (-i k \hat{x} -\pi^{++a} \hat{\theta}^{++}_{a} -\pi^{--a} \hat{\theta}^{--}_{a} -\pi^{0a} \hat{\theta}^{0}_{a}), \nonumber \\ 
 dp&=& d^3 k d^2 \pi^{++} d^2 \pi^{--} d^2 \pi^{0}, \nonumber \\ 
V^{++} (p) &=& V^{++} (k, \pi^{++},\pi^{--},\pi^{0}, u^{\pm}  ).
\end{eqnarray}
Thus,  we  obtain  a one to one map between a function of
$\hat z$ to a function of ordinary
 superspace coordinates $z$ via
\begin{equation}
V^{++} (z)  =
\int dp V^{++}(p) \exp ( i p z).
\end{equation}
 We can express the product of two fields  
${\hat V^{++}}(\hat z) { \hat V^{++}}
 (\hat z )$
on this deformed superspace as
\begin{eqnarray}
{\hat V^{++}}( \hat z ) { \hat V^{++}}  ( \hat z ) 
=
\int dp_1 dp_2
\exp i( ( p_1 + p_2) \hat z ) \exp(i\Delta) V^{++} (p_1) V^{++}(p_2), &&
\end{eqnarray}
where
\begin{equation}
\exp (i\Delta) = \exp -\frac{1}{2} \left(
C^{++ ab} \theta^{++ 2}_a \theta^{++1}_b \right).
\end{equation}
This motivates the  definition of  the star product
  between ordinary vector fields, which is now defined as   
\begin{eqnarray}
V^{++} (z) \star V^{++}  (z) &=&\exp -\frac{1}{2} \left(
C^{++ ab} \partial^{ ++2}_a \partial^{++1}_b \right) \nonumber \\  && \times
 {V^{++}}(z_1) { V^{++}}  (z_2)
\left. \right|_{z_1=z_2=z}. 
\end{eqnarray}
 If we impose the following deformation 
\begin{equation}
 \{\hat \theta^{++}_a,\hat  \theta^{++}_a \} = C^{++}_{ab}, \quad \{\hat \theta^{--}_a, \hat \theta^{--}_a \} = C^{--}_{ab},
\end{equation}
and proceed in a similar way, we obtain the following definition of  the star product
  between ordinary vector fields
\begin{eqnarray}
V^{++} (z) \star V^{++}  (z) &=&\exp -\frac{1}{2} \left(
C^{++ ab} \partial^{ ++2}_a \partial^{++1}_b + C^{-- ab} \partial^{ --2}_a \partial^{--1}_b\right)
\nonumber \\ && \times 
 {V^{++}}(z_1) { V^{++}}  (z_2)
\left. \right|_{z_1=z_2=z}. 
\end{eqnarray}
However, we will only analyse the deformation corresponding to Eq. (\ref{p}) in this paper. 
So, in this paper the 
deformed ABJ model will have manifest $\mathcal{N} = 2$ supersymmetry. 
It is now possible to write $V^{--}$ in-terms of $V^{++}$ as
\begin{equation}
 V^{--}(z,u)=\sum_{n=1}^\infty (-1)^n \int du_1\ldots
du_n  E^{++},
\end{equation}
where 
\begin{equation}
 E^{++} = \frac{V^{++}(z,u_1)\star V^{++}(z,u_2)\ldots \star 
V^{++}(z,u_n)}{(u^+u^+_1)(u^+_1u^+_2)\ldots (u^+_n u^+)}.
\end{equation}

Now we can write the action for the deformed ABJ theory. This theory is invariant under the gauge group $U(N) \times U(M)$.
In this theory the matter fields are denoted by $(q^{+})_A^{\underline B}$ and $(\bar{q}^+)^A_{\underline B}$  and the gauge 
fields corresponding to $U(M)$ and $U(N)$ are denoted by  $(V^{++}_L)^A_B$ and
$(V^{++}_R)^{\underline{A}}_{\underline{B}}$, respectively. 
 Here the  underlined indices refer
to the right $U(M)$ gauge group. The covariant derivatives for the matter fields in the deformed ABJ theory can be written as  
\begin{eqnarray}
\nabla^{++}q^{+}&=&{  D}^{++}q^{+}
 + V^{++}_L \star q^{+}- q^{+} \star V^{++}_R\,,  \nonumber \\   \nabla^{++}\bar q^{+}&=&{  D}^{++}\bar q^{+}
 -\bar q^{+}  \star V^{++}_L +  V^{++}_R  \star \bar q^{+}\, ,
\end{eqnarray}
We can write the action for the ABJ theory in the deformed harmonic superspace as, 
\begin{equation}
 S = S_{CS, k} [ V^{++}_L]_ \star  + S_{CS, - k} [ V^{++}_R]_ \star   + S_{M} [ q^{+}, \bar q^{+}]_ \star,
\end{equation}
where 
\begin{eqnarray}
 S_{CS, k}[ V^{++}_L]_ \star &=&\frac{ik}{4\pi}\, tr\sum\limits^{\infty}_{n=2} \frac{(-1)^{n}}{n} \int
d^3x d^6\theta du_{1}\ldots du_n  H^{++}_L, \nonumber \\
  S_{CS, -k}[ V^{++}_R]_ \star &=&- \frac{ik}{4\pi}\,tr\sum\limits^{\infty}_{n=2} \frac{(-1)^{n}}{n} \int
d^3x d^6\theta du_{1}\ldots du_n H^{++}_R, \nonumber \\
S_{M} [ q^{+}, \bar q^{+}]_ \star &=&tr\int d^3 x d\zeta^{(-4)}\bar q^{+} \star \nabla^{++}  \star q^{+},
\end{eqnarray}
where 
\begin{eqnarray}
 H^{++}_L &=& \frac{V^{++}(z,u_{1} )_L  \star V^{++}(z,u_{2} )_L\ldots
 \star V^{++}(z,u_n )_L }{ (u^+_{1} u^+_{2})\ldots (u^+_n u^+_{1} )},
\nonumber \\ 
H^{++}_R &=& \frac{V^{++}(z,u_{1} )_R  \star V^{++}(z,u_{2} )_R\ldots
 \star V^{++}(z,u_n )_R }{ (u^+_{1} u^+_{2})\ldots (u^+_n u^+_{1} )}.
\end{eqnarray}
This theory is invariant under the following  infinitesimal gauge transformations
\begin{eqnarray}
\delta q^{+} &=& \Lambda_L  \star q^{+}-q^{+} \star \Lambda_R,\nonumber \\
 \delta\bar q^{+} &=&\Lambda_R  \star \bar q^{+}-\bar q^{+} \star \Lambda_L,\nonumber \\
\delta V^{++}_L&=&-{  D}^{++}\Lambda_L -[V^{++}_L,\Lambda_L]_ \star,\nonumber \\
\delta V^{++}_R&=&-{  D}^{++}\Lambda_R -[\Lambda_R, V^{++}_R]_ \star.
\end{eqnarray}
 The  deformation of the ABJ theory breaks the 
supersymmetry  from $\mathcal{N} =3$ supersymmetry to $\mathcal{N}= 2$ supersymmetry. However, the original ABJ theory had $\mathcal{N}  = 6$
supersymmetry. We have broken manifest the supersymmetry corresponding to $Q^{++}_a$, so we should still be left with $\mathcal{N} =5$ supersymmetry. 
Now as we have manifest $\mathcal{N} =2$ supersymmetry generated by to the supercharges $Q^{--}_a$ and $Q^{0}_a$, 
 we should have an additional $\mathcal{N} =3$ supersymmetry to generate $\mathcal{N} =5$ supersymmetry.  This is achieved by the following 
supersymmetric transformations, 
\begin{eqnarray}
\delta_\epsilon q^{+}&=& i\epsilon^{a}\hat\nabla^0_a \star q^{+}\,, \nonumber \\
\delta_\epsilon\bar q^{+} &=&i\epsilon^{a} \hat\nabla^0_a \star \bar q^{+ }\,, \nonumber  \\
\delta_\epsilon V^{++}_L&=&\frac{8\pi}k\epsilon^{a}
 \theta^0_a \star q^+\bar\star q^+\,, \nonumber \\
\delta_\epsilon V^{++}_R &=&\frac{8\pi}k\epsilon^{a}
 \theta^0_a \star \bar q^+ \star q^+\,,
\label{epsilon4}
\end{eqnarray}
where
\begin{eqnarray}
 \hat\nabla^0_a \star q^{+}& =& \nabla^0_a \star q^{+} 
+\theta^{--}_a (W^{++}_L \star q^{+} -q^{+} \star W^{++}_R )\,, \nonumber \\
 \nabla^0_a \star q^+&=&D^0_a q^+
 +V^0_{L\, a}\star q^+ -q^+\star V^0_{R\,a }\,, \nonumber \\ V^0_{L\, a}&=&-\frac12D^{++}_a
V^{--}_{L}, \nonumber \\ 
V^0_{R\, a}&=&-\frac12D^{++}_a
V^{--}_{R}.
\end{eqnarray}
Here $\hat\nabla^0_a \bar q^{+ } $ and $  \nabla^0_a \bar q^{+ }$ are obtained via  conjugation and
the field strengths $W^{++}_R$ and $W^{++}_L$ are defined by 
\begin{eqnarray}
 W^{++}_L &=& -\frac{1}{4} D^{++a} D^{++}_{ a}  V^{--}_L, \nonumber \\ 
 W^{++}_R &=& -\frac{1}{4} D^{++a} D^{++}_{ a}  V^{--}_R. 
\end{eqnarray}
These field strengths satisfies 
\begin{eqnarray}
 D^{++} W^{++}_L  + [V^{++}_L , W^{++}_L ]_\star &=0, \nonumber \\ 
 D^{++} W^{++}_R + [V^{++}_R, W^{++}_R]_\star &=0.
\end{eqnarray}
Now using Fierz rearrangement, we get $-   \delta_\epsilon 
S_{M} [ q^{+}, \bar q^{+}]_ \star =  \delta_\epsilon S_{CS, k} [ V^{++}_L]_ \star  +  \delta_\epsilon S_{CS, - k} [ V^{++}_R]_ \star $,
and so we have 
\begin{equation}
 \delta_\epsilon  S =0,
\end{equation}
and thus the action is invariant under $\mathcal{N} =5$ supersymmetry. Thus, unlike the undeformed ABJ theory which is
 invariant under $\mathcal{N} =5$ supersymmetry,
the deformed ABJ theory is only invariant under $\mathcal{N} =5$ supersymmetry.

\section{Conclusion}

In this paper we analysed the non-anticommutative deformation of
 the ABJ theory in harmonic superspace. This theory is  dual  
to multiple D2-brane in $RR$ background. 
The full multiple D2-brane action includes couplings 
to the background fields of type II string theory.
For a single brane this would be the pull back of $C_{\mu\nu\tau}$
 to the world volume but for the non-Abelian multiple
D2-brane action it must include further dielectric couplings to all of the $RR$ form fields.
 M-theory contains a background three 
form field and its dual field. These  should reduce to
the $RR$ to fields of string theory. 
The D-branes have been studied in various backgrounds. 
 So, it will be interesting to analyse the coupling of the three form  
field explicitly  to the multiple M2-brane
action in  harmonic superspace. 
 
As the $RR$ background partially break the supersymmetry in type II string theory, 
the dual deformations of it will also partially break the supersymmetry on the M-theory side. 
We analyse the M2-branes in harmonic superspace. 
This  harmonic superspace initially had manifest $\mathcal{N} =3$ supersymmetry. 
However, after the non-anticommutative deformation, it only had 
$\mathcal{N} =2$ supersymmetry. 
This was because the supersymmetry corresponding to $Q^{++}_a$ was broken by the imposition of the 
non-anticommutative deformation of the superspace. Thus, the total supersymmetry of the ABJ theory was reduced from $\mathcal{N} = 6$
to $\mathcal{N} =5$ supersymmetry. We also discussed the deformations of the harmonic superspace that break the supersymmetry corresponding 
to two of the supercharges, namely $Q^{++}_a$ and $Q^{--}_a$, respectively.
 This deformation only leaves manifest $\mathcal{N} = 1$ supersymmetry
unbroken. Thus a similar analysis of the ABJ theory in this superspace 
will only have $\mathcal{N} = 4 $ supersymmetry. 
There are other type of deformations that occur in superspace. These 
occur due to non-vanishing values of commutators between spacetime and 
Grassmann coordinates and physically correspond to a deformation generated by a gravitino background. It will be interesting to study the ABJ theory 
with this kind of deformations in harmonic superspace. The interesting thing about these deformations is that they do not break any amount of supersymmetry. 
Thus, the ABJ theory with deformed superspace, where the deformations are 
caused by a gravitino background will preserve all of the 
$\mathcal{N} =6$ supersymmetry.

Chern-Simons-matter theories also  have important  applications in  condensed matter physics. 
This is because of their  relevance 
 to the fractional quantum Hall effect, which is based 
on the concept of statistical transmutation. 
\cite{aaaaa, baaaa, caaaa, daaaa}. 
Recently,  supersymmetric 
generalisation of the fractional quantum Hall effect has also
 been investigated \cite{a11, all1, all2, all3}.
In particular, physical properties of the topological
excitations in the supersymmetric quantum Hall liquid were discussed in a dual supersymmetric
Chern-Simons theory \cite{a5all}.  Furthermore,  a close connection between 
the fractional quantum Hall 
  noncommutativity  of the spacetime has been discovered
\cite{fqhfqh1,fqhfqh2a, fqhfqh2b, fqhfqh2}. 
Thus, the results of this paper  can have interesting  condensed matter applications. 
This is because we can analyse the non-anticommutative deformation 
of the supersymmetric fractional quantum Hall effect. This can
  change the behavior  of  fractional condensates and thus
have  important consequences for the transport properties in  
supersymmetric  quantum hall  systems.
  
 It may be noted that the BRST symmetry of the ABJM theory 
has been analysed in deformed $\mathcal{N} =1$ superspace 
\cite{3, 4, qwer}. So, it will be interesting to
 analyse the BRST symmetry of ABJ theory 
in deformed harmonic superspace. We can also 
use the BRST symmetry of 
this theory  to show the unitarity of the $S$-matrix. 
It is possible to reduce the ABJM action to a $\mathcal{N} =8$,  
super-Yang-Mills theory describing N
D2-branes by using the novel Higgsing mechanism \cite{zz1,zz2,zz3,zz4a}. In this Higgsing 
mechanism the gauge group of the ABJM theory is spontaneously broken down to its 
diagonal subgroup. This analysis has also been performed in $\mathcal{N} =1$ superspace \cite{4,hi}, 
 and 
it will interesting to repeat this analysis in harmonic superspace. By doing that we will be able 
to analyse this Higgsing mechanics for the ABJM theory 
with non-anticommutative deformations. This will give us a better understanding of 
the existence of these non-anticommutativite deformation in the M-theory, 
as we will be able to relate it to the familiar objects in the 
string theory.

\end{document}